\documentstyle[11pt,paspconf,epsf]{article}

\def\etal{et al.}
\def\hii{H{\sc ii}}
\def\cbeta{$c_{\rm H\beta}$}

\def\msun{M$_{\odot}$}
\def\zsun{Z$_{\odot}$}
\def\halpha{\ifmmode {\rm H{\alpha}} \else $\rm H{\alpha}$\fi}
\def\hbeta{\ifmmode {\rm H{\beta}} \else $\rm H{\beta}$\fi}
\def\heiia{He\,{\sc ii} $\lambda$4686}

\def\niii{N\,{\sc iii} $\lambda$4640}

\def\Ciii{C\,{\sc iii} $\lambda$4650}
\def\ciii{C\,{\sc iii} $\lambda$5696}
\def\civ{C\,{\sc iv} $\lambda$5808}
\begin{document}

\title{Populations of WC and WN stars in WR galaxies: new constraints 
on starburst parameters}
\author{Thierry Contini}
\affil{European Southern Observatory, Garching, Germany}
\author{Daniel Schaerer}
\affil{ UMR 5572, Observatoire Midi-Pyr\'en\'ees, Toulouse, France}
\author{Daniel Kunth}
\affil{Institut d'Astrophysique de Paris, France}

\begin{abstract}

Detection and analysis of Wolf-Rayet (WR) star populations in young 
starbursts is a powerful tool to reveal the properties of the 
star-forming events. We present new high S/N optical spectra of 
a sample of well-known WR galaxies, in which we detected for the 
first time the spectral signature of WC stars. The detailed comparison 
of WN and WC star populations derived in these galaxies, with the 
most recent model predictions, allows to put tight constraints 
on starburst parameters, such as the age and duration of the burts, 
and the IMF.  

\end{abstract}

\section{Introduction}
Starbursts play a major role in the global process of galaxy formation and 
evolution. A better knowledge of their intrinsic properties (duration, IMF, 
etc) in nearby starburst 
galaxies is a key issue to follow the evolution of the large 
population of star-forming galaxies observed up to redshift $z \sim 4$ (e.g. 
Steidel et al. 1998).
The so-called ``Wolf-Rayet (WR) galaxies'' may be the ideal laboratories 
to derive starburst properties since these objects harbour the most massive 
stars known, O stars and their descendents WR stars, which allow to 
probe the upper part of the IMF and the youngest stellar populations.
    
Since the initial compilation of Conti (1991) listing 37 galaxies, 
the number of known WR galaxies has grown rapidly to about 130 in 
the last compilation of Schaerer, Contini \& Pindao (1999, hereafter SCP99). 
While most of them fall in the category of ``classical'' star-forming 
galaxies (\hii\ galaxies or Starburst Nucleus Galaxies), broad WR 
emission lines have also been detected in more ``exotic'' objects 
like Seyfert 2 galaxies (e.g.\ Heckman \etal\ 1997; Kunth \& Contini, 
these proceedings) or giant cD galaxies, 
located in the centre of galaxy clusters with strong cooling flows 
(Allen 1995).

The presence of a large number of WR stars has already provided interesting 
informations about star formation in WR galaxies, indicating 
recent ($\la 10$ Myr) star-forming events and the existence of 
massive stars (e.g.\ Vacca \& Conti 1992, hereafter VC92). 
But in practice, the detection of the broad emission line \heiia, 
attributed mainly to WR stars of WN subtype, was often simply used.
The strongest emission line of WC stars, \civ, has quite rarely been 
observed so far in WR galaxies. 
Although Conti (1991) claimed the absence of convincing evidence for 
carbon features in WR galaxies, there is 
no doubt about their existence anymore since about 56 objects showing 
both broad \heiia\ and \civ\ features are listed in SCP99.

The existence of WC stars in WR galaxies, although possibly less numerous 
than WN stars, 
is indeed expected both from observations of WR populations in the Local 
Group (e.g.\ Massey \& Johnson 1998, hereafter MJ98), and from stellar 
evolution models (Maeder \& Meynet 1994). 
Surprisingly, the homogeneous, fairly high S/N spectra from 
the sample of VC92 containing 12 regions with \heiia\ detections 
revealed only one \civ\ detection (in He 2-10 A). 
If true, this would certainly contradict the predictions from synthesis
models for young starbursts (Meynet 1995, Schaerer \& Vacca 1998, 
hereafter SV98), which otherwise compare well to observations in the 
Local Group (Maeder \& Meynet 1994).
To verify if this apparent discrepancy really holds we have searched 
for WC stars in well-known WR galaxies (NGC~3049, He 2-10, NGC~3125, 
NGC~5253, Tol~89). 
Previous accounts of our results are given in Schaerer et al. (1997) 
and Schaerer, Contini \& Kunth (1999, hereafter SCK99). 


\section{Observations and spatial analysis}
\label{SPATA}

Long-slit spectra of galaxies were obtained on the nights of 1995 April 24 -- 
26 at the ESO 2.2m telescope. The data were acquired with the EFOSC2 
spectrograph which gives a spectral coverage of 4400 -- 6500 
\AA\ with a resolution of $\sim$ 5 \AA. The slit was oriented in order to 
cover regions with previous 
detections of WR stars. 
The total integration time for each galaxy ranges from 80 to 120 min. 
The seeing was relatively 
stable during the observations with a mean spatial resolution of about 
1\arcsec. 
The procedure for 
data reduction is fully described in SCK99.

We used our long-slit spectroscopic observations to compare, 
for each galaxy, 
the spatial distributions of nebular and WR emission-line intensities 
and the stellar continuum emission.
The results for each galaxy are given in 
SCK99; they can be summarized as follows. There is an offset between the 
distribution of nebular lines and stellar continuum in He 2-10 and Tol 89.
The WR emission features follow the stellar continuum with no 
significant shift between \heiia\ and \civ. The spatial extension of WR 
features ranges from $\sim$ 60 pc (NGC 5253) to $\sim$ 270 pc (He 2-10 and 
NGC 5253). Two distincts WR regions are identified in NGC 3125 and NGC 5253.


\section{The WR regions}

\begin{figure}
\plottwo{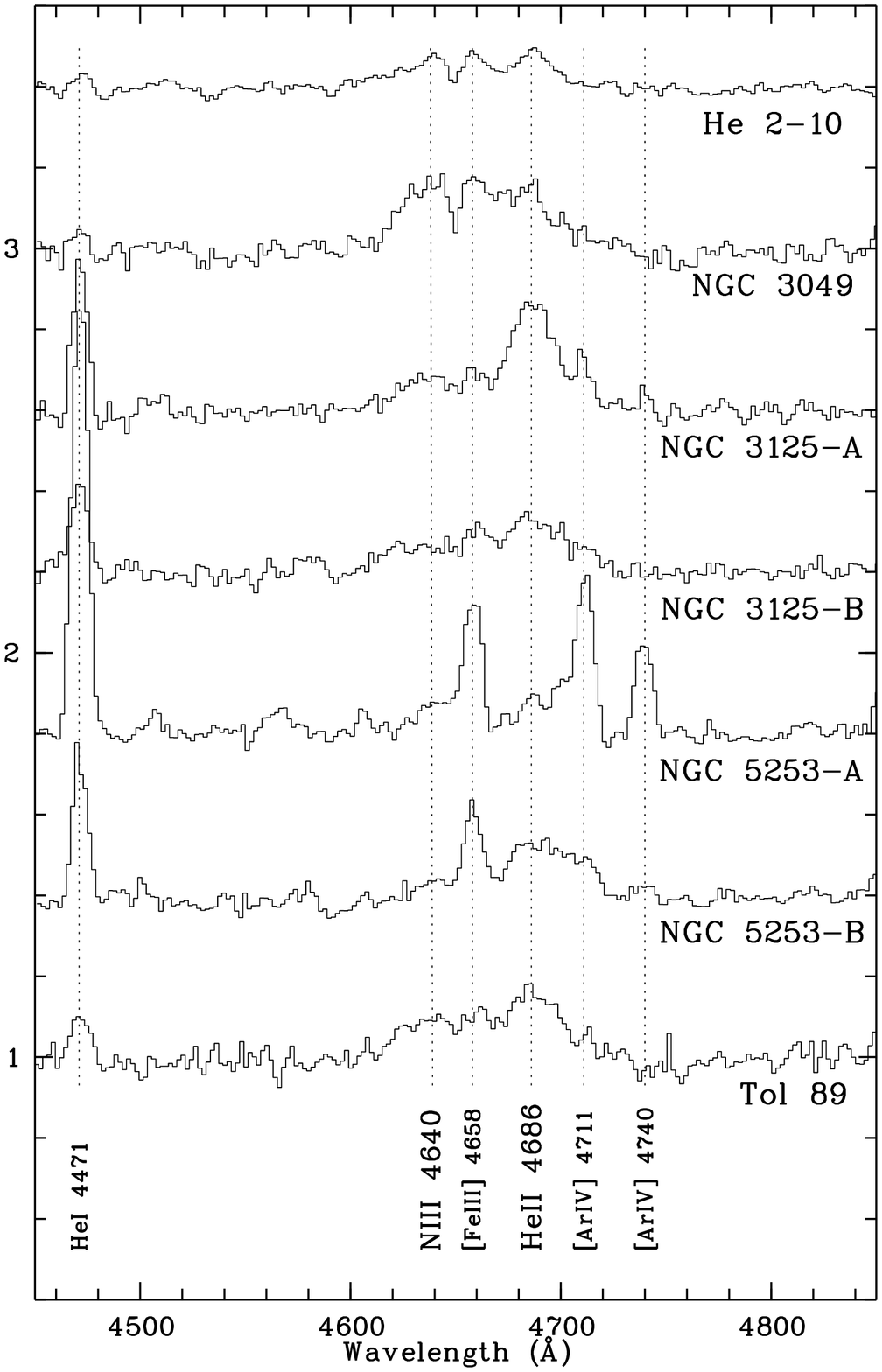}{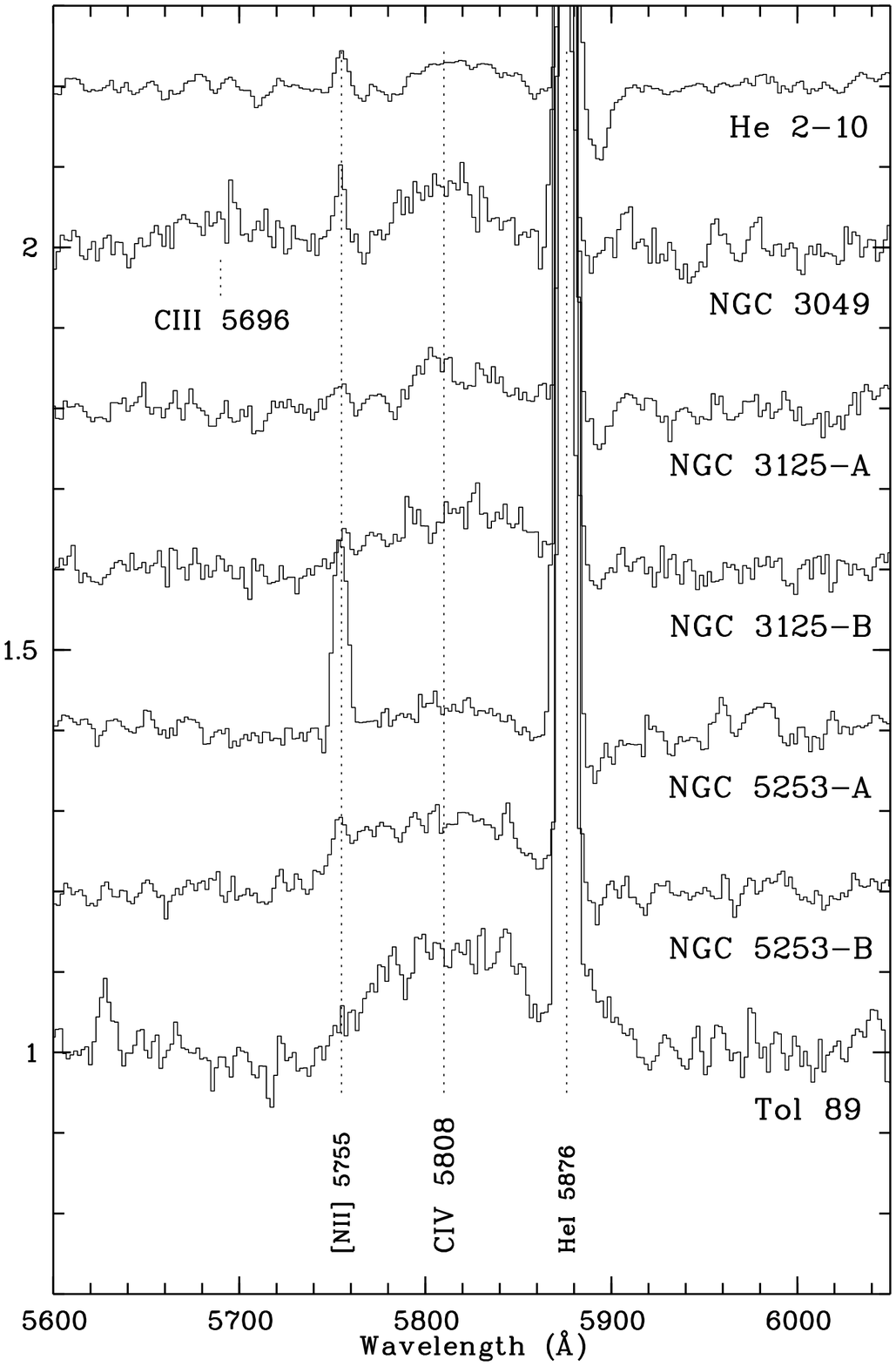}
\caption{Spectra of galaxies showing the ``blue'' (left) and 
``red'' (right) WR bumps}
\label{SPEC}
\end{figure}

One-dimensional spectra were extracted corresponding to the different
regions where spectral signatures from WR stars have been detected.
Normalized spectra showing the region around the ``blue'' and ``red'' 
WR bumps are given in Fig.~\ref{SPEC}. 
Several broad emission lines due to WR stars have been 
detected over the wavelength range 
covered by our spectra: the blend of 
\niii\ and \Ciii, \heiia, and \civ.
Equivalent widths (EWs) of these emission lines are listed in 
Table~\ref{DATA}, together with the basic nebular parameters 
(metallicity, extinction, etc). 

One of the main results of these observations is the unambiguous detection of  
broad (FWHM  $\sim 50 - 90$ \AA) \civ\ emission in all WR regions of 
the observed galaxies (see right panel of Fig.~\ref{SPEC}), 
which clearly indicates {\em the presence of WC stars in these regions}.
Whereas all the objects were previously known to show the broad WR bump around 
4700 \AA, our spectra are the first to show the red WR bump around 5800 \AA.
As already noted (Schaerer et al. 1997, SCK99), a sufficiently high S/N 
($\ga$ 40) is required 
for the detection of such a \civ\ line in integrated spectra of WR galaxies.
This explains why this line has remained undetected in previous ground-based
spectroscopical studies, except for He~2-10 (VC92). 

In all our seven emission-line regions we thus find signatures of
both WN and WC stars. The dominant WR subtypes in each region can be constrained 
using measured line widths and intensity ratios between WR emission lines (see 
SCK99 for a full description). All regions (except the high-metallicity galaxy 
NGC~3049) contain a mixed population of WNL, and early WC 
and/or WO3-4 stars. In NGC~3049, the detection of \ciii\ indicates the 
predominance of WCL stars. The presence of intermediate-type WN/WC stars 
cannot be excluded.

\begin{table}
\caption{The broad WR emission lines}
{\footnotesize
\begin{tabular}{lrrrrrrr}
\hline
\hline
Line/Galaxy    & He~2-10 &  NGC~3049 & \multicolumn{2}{c}{ NGC~3125} & \multicolumn{2}{c}{NGC~5253} & {Tol~89}  \\
\cline{4-5}
\cline{6-7}
        &         &          &   A & B                       & A & B                        &        \\
\hline 
$EW$(\hbeta) [\AA]&	    23$\pm$ 3&	    44$\pm$ 3&	   93$\pm$  9&	   70$\pm$  7&	  269$\pm$ 23&	   112$\pm$ 10&	   68$\pm$ 17\\
(O/H) [\zsun]	 &  0.42 &  1.20 &  0.17 & 0.20 & 0.20 & 0.20 & 0.25 \\
\cbeta\		 &0.56&0.23&0.40&0.64&0.44&0.20&0.18\\
\hline
 & \multicolumn{7}{c}{\sc  EW [\AA] of broad emission lines from Wolf-Rayet stars } \\[0.1cm]
\hline
\niii &	  2.7$\pm$0.5&    5.2$\pm$0.5&	  3.2$\pm$0.5&	  2.8$\pm$1.0&	  2.1$\pm$1.3&	  1.0$\pm$0.5&	  3.4$\pm$0.5\\
+\Ciii &	 &    &	 & &&&\\
\heiia &	  2.3$\pm$0.5&	  5.6$\pm$0.5&	  7.3$\pm$1.0&	  2.8$\pm$0.5&	  3.0$\pm$0.7&	  3.3$\pm$1.1&	  6.1$\pm$0.5\\
\ciii &\ldots	     &	  2.0$\pm$0.5&\ldots	     &\ldots	     &\ldots	     &\ldots	     &\ldots	     \\
\civ	&	  1.5$\pm$0.5&	  4.7$\pm$0.6&	  3.1$\pm$0.4&	  6.2$\pm$1.0&	  2.5$\pm$0.6&	  6.0$\pm$1.1&	  11.9$\pm$1.5 \\
\hline
\hline
\end{tabular}
}
\label{DATA}
\end{table}

\section{Populations of WR and O stars}

The approximate number of WN, WC, and O stars derived from the observed 
emission lines are listed in Table~\ref{STARPOP}. 
The number of WN and WC  stars is calculated from the luminosity of the WR 
emission lines, assuming that the dominant contributors to the broad 
\heiia\ and \civ\ lines are respectively WNL and WC4 stars 
(see previous section). 
Under the condition of case B recombination and assuming that all the 
ionizing photons emitted by the stars are absorbed by the gas, the total 
number of Lyman photons can be derived from the observed 
luminosity of the \hbeta\ emission line. 
The number of O stars is thus estimated, taking into account 
the ionizing photon contribution from WR stars, the age of the stellar population 
and the IMF (see SCK99). 
The number of O stars found for the WR regions are between $\sim$ 500 and 7000. 
The case of NCG 3049 is somewhat exceptional (see SCK99 for a discussion). 
According to the models, the bulk of the ionisation in this galaxy is provided 
by late O and/or B type stars and WR stars.
Uncertainties on absolute number of WR and O stars reported in 
Table~\ref{STARPOP} are quite large. 
The main sources are the following (see SCK99 for details). There is a 
large uncertainty on the average luminosity of WNL stars in the \heiia\ 
line, and to a lesser extent of WCE stars in the \civ\ line. 
The dominant WC subtype is not well constrained. 
There may be a nebular contribution to the \heiia\ emission in NGC 3125-A 
and NGC 5253-A (see Fig~\ref{SPEC}). A significant fraction of ionizing 
photons may escape from \hii\ regions. 


The number ratios WR/O and WC/WN are reported in 
Table~\ref{STARPOP}. We obtain WR/O number ratios of $\sim 0.03 - 0.4$
(and even larger for NGC~3049), systematically higher than the predictions 
for constant star formation at the appropriate metallicity (Maeder \& Meynet 1994), 
but within the range of instantaneous
burst models with different IMF slopes (SV98).
A trend of increasing WR/O ratios towards higher metallicity is found
as expected (e.g.\ Meynet 1995).
The number ratio of WC/WN stars is typically $0.2 - 0.4$, with no systematic 
variation with metallicity. 
For the low-metallicity objects ($Z \sim$ 0.2 \zsun; NGC 3125, NGC~5253, and 
Tol~89) the derived WC/WN ratio is larger than what is found in the Local Group 
(except IC~10) at similar metallicity (MJ98). 
At higher metallicities, He~2-10 and NGC~3049 show, however, WC/WN ratios below the 
trend observed by MJ98.

The finding of a fairly constant WC/WN ratio may seem surprising at first sight.
Contrary to regions of constant star formation, likely representative of the Local
Group samples, regions of short star formation could a priori show quite a large range 
in WC/WN, depending on the age of the starburst. 
However, as explained in SCK99, burst durations of $\ga 2$ Myr already suffice to 
smooth out the WC/WN ratio. This may well explain the small range of 
WC/WN found for the low-metallicity objects. 
More surprising is the low WC/WN ratio of the remaining higher metallicity objects 
(He~2-10 and NGC~3049) compared to the WC/WN value of MJ98 at a similar metallicity. Indeed, the probability of finding WC/WN ratios below 
the equilibrium value attained in regions of constant star formation should be quite 
small.
However, if the number of WC stars is systematically underestimated by a factor of 3, 
our observations may well all be larger than the observed WC/WN trend with
$Z$.
We conclude that the WC/WN ratios of the low-metallicity galaxies 
can be understood quantitatively with burst models of reasonably short 
but non-zero duration. 
Additional observations of WC and WN populations, especially for regions of higher 
metallicities, would be very helpful.

\begin{table}
\caption{Populations of WR and O stars in WR regions}
{\footnotesize
\begin{tabular}{llcrrcccc}
\hline
\hline
Galaxy &  & Age burst & WN & WC & O &  WR/O &  WC/WN  \\
       &  &[Myr]      &    &    &   &      &       \\
\hline
 He 2-10  &  & 5.5 -- 6.0 & 1100$\pm$520 & $>$ 250  & 2450 -- 4900 & 0.37 & 0.23 \\
 NGC 3049  &  & 5.5        &  510$\pm$240 & $>$ 170  & ...           & ...   & 0.33 \\
 NGC 3125 & A & 4.5 -- 5.0 &  500$\pm$230 & $>$  70  & 3240 -- 6470 & 0.12 & 0.14 \\
         &  B & 4.5 -- 5.0 &  530$\pm$250 & $>$ 200  & 3450 -- 6900 & 0.14 & 0.38 \\
 NGC 5253 & A &    3.0     &   26$\pm$13  & $>$   9  & 960  -- 1080 & 0.03 & 0.35 \\
         &  B & 5.0        &   27$\pm$13  & $>$  10  & 680          & 0.05 & 0.37 \\
 Tol 89   &  & 4.5 -- 5.0 &  240$\pm$110 & $>$ 150  & 640  -- 1270 &  0.41 & 0.62  \\
\hline
\hline
\end{tabular}
}
\label{STARPOP}
\end{table}

\section{Age, duration and IMF of starbursts}

One can try to constrain the main burst parameters
(age, IMF, star formation history) by comparing the observed 
WR signatures to the predictions from the latest population 
synthesis models for young starbursts (SV98). 
We consider two basic free model parameters: 
the IMF slope and the duration of the star-forming event.
The metallicity $Z$ and the starburst age (from $EW$(\hbeta)) are adopted 
from the observations.
A power law with a slope $\alpha$ is adopted for the IMF 
(in our notation $\alpha = 2.35$ for a Salpeter IMF). 
The upper mass cut-off is generally set to 
$M_{\rm up}=$ 120 \msun.
We consider burst models with different durations $\Delta t$ starting 
at time $t=0$, with the limiting case of an instantaneous burst
($\Delta t=0$). SV98 models are based on the latest Geneva stellar evolution tracks 
and only single-star models are considered. A detailed discussion on the 
assumptions or uncertainties of the models is given in SCK99.

The comparison with the observations are performed in two ways.
One both compare relative line intensities of the WR features with
respect to the nebular \hbeta\ emission (in short WR/\hbeta), and 
equivalent widths of the WR lines. 
Observationaly the determination of these quantities can mainly be ``perturbed'' 
by three effects:
\begin{itemize}
\item[{\em (i)}] 
We do not count all the ionizing photons produced by massive stars.
  This may happen if the nebular emission from the \hii\ region of interest
  is not entirely included in the slit, and/or if
  ionizing photons can escape from the \hii\ region.
\item[{\em (ii)}] Stars and gas suffer from a different extinction.
\item[{\em (iii)}] An underlying older population contributes additional 
	continuum light.
\end{itemize}
The relative intensities WR/\hbeta\ are affected only by  {\em (i)} and 
{\em (ii)}.
Effect {\em (i)} increases the relative WR/\hbeta\ intensities.
The same holds for  {\em (ii)} if the stellar light is less extincted
than the gas (cf.\ e.g.\ Calzetti 1997, Mas-Hesse \& Kunth 1999).
The EWs are only affected by {\em (iii)}, which decreases 
the observed value. We refer to SCK99 for a discussion on the quantitative 
importance of these effects.
Given these potential difficulties both comparisons of relative line 
intensities
WR/\hbeta\ and WR equivalent widths have been performed for all objects. 
Significant differences between the two methods likely indicate some difficulty
with {\em (i)} and {\em (iii)}.
In this case the simplest meaningful comparison is between the observed 
WR equivalent widths and the maximum value predicted by the models 
(irrespectively of the age); it only depends on {\em (iii)}.

\begin{figure}
\plottwo{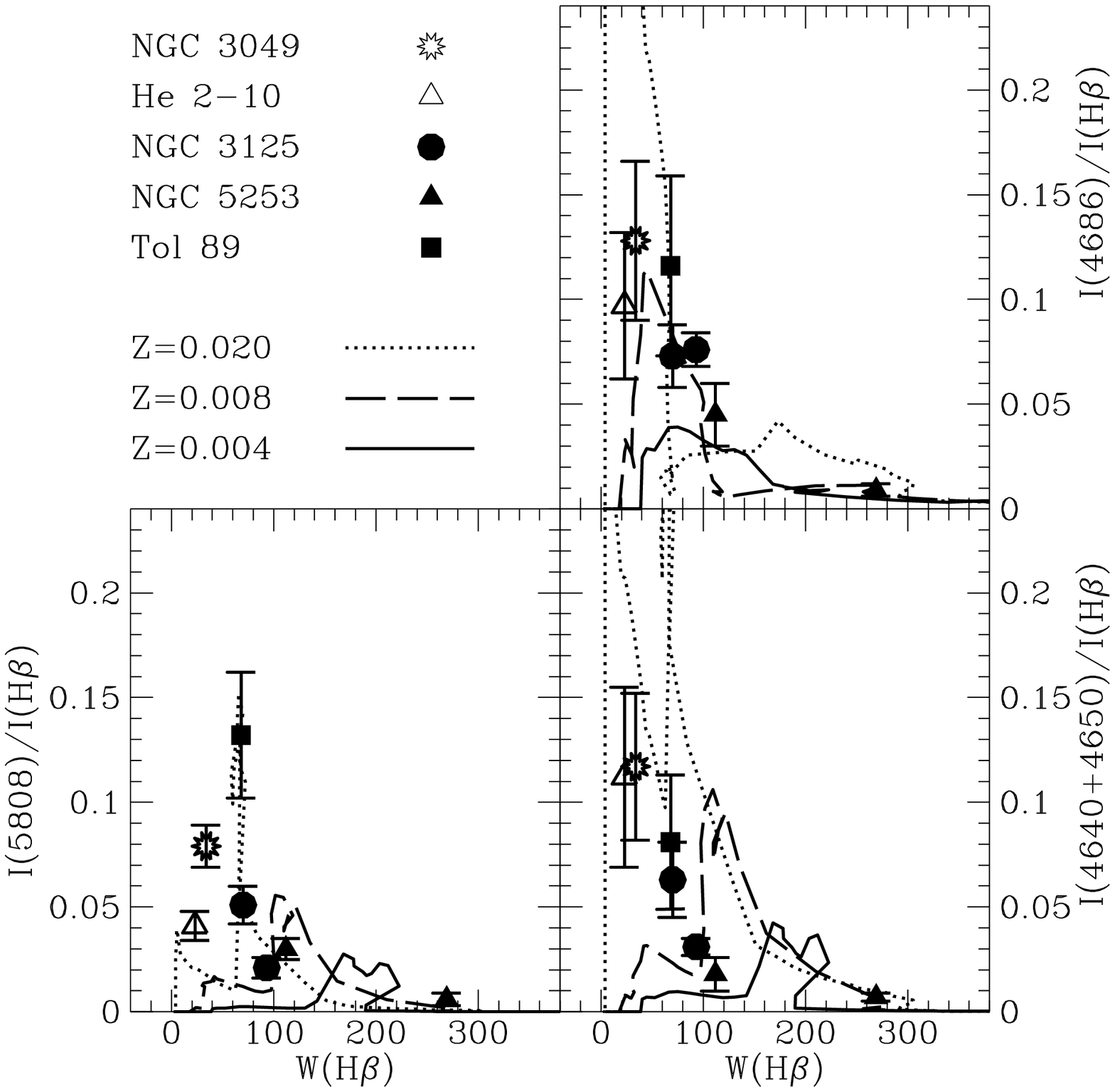}{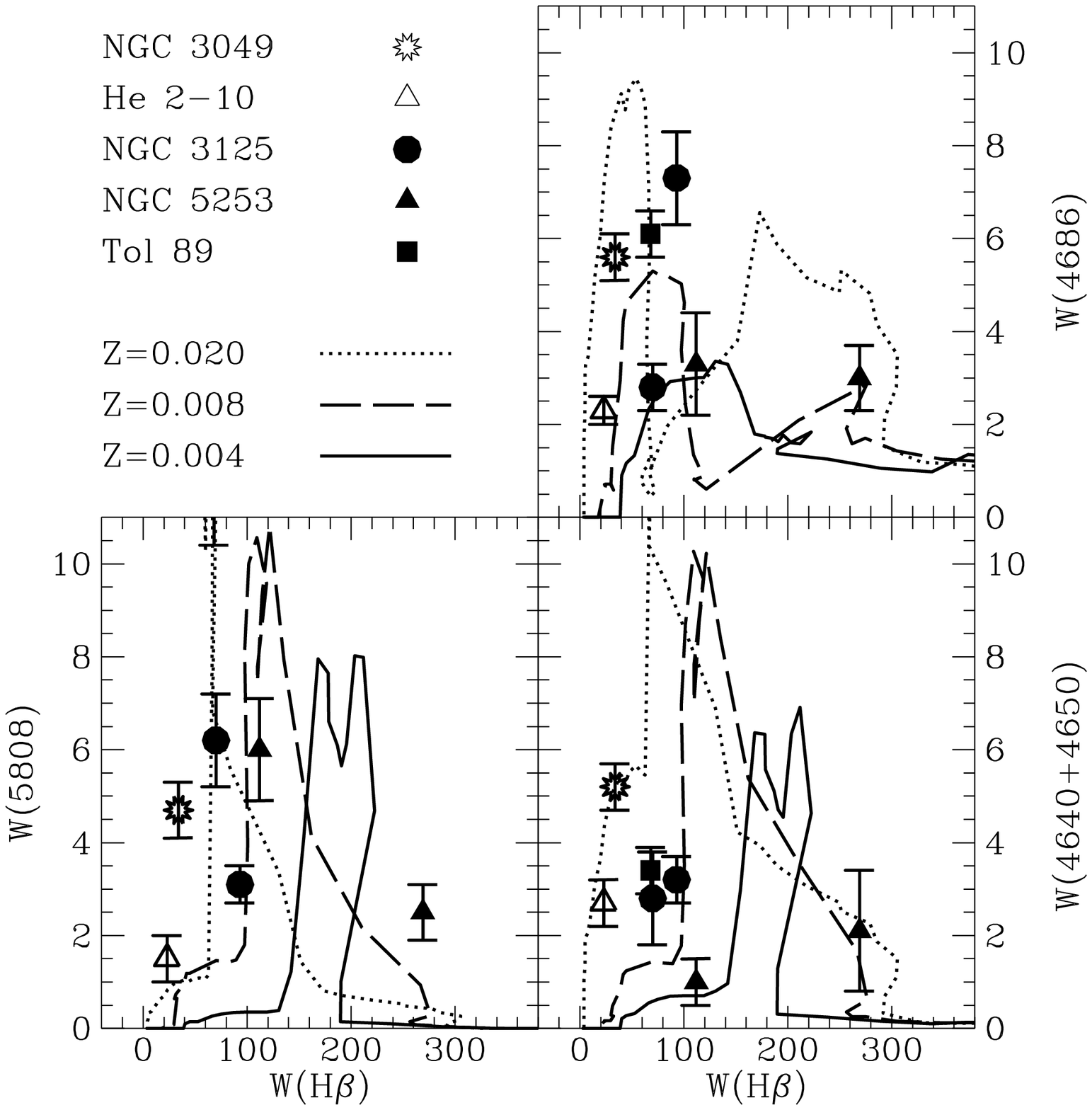}
\caption{Observed vs. predicted WR line intensities relative to 
\hbeta\ (left) and WR equivalent widths (right)}
\label{MODELS}
\end{figure}

In Fig.~\ref{MODELS} we compare, for all the WR regions, the observed 
relative intensities and EWs of WR lines with the model 
predictions of SV98 at metallicities $Z=$ 0.02 (solar), 0.008, and 0.004 
for an instantaneous burst with a Salpeter IMF. In both plots, $W(\hbeta)$ 
is used as a time indicator (decreasing $W$ with time) in the 
synthesis models.
Comparison of the individual objects (classified by metallicity) with the 
model predictions can be summarized as follows.

\subsection{High-metallicity galaxies}

The predictions of the standard model (instantaneous burst, 
Salpeter IMF) are in agreement with the observed WR line strengths 
of the high-metallicity galaxies, particularly for NGC 3049.
For He 2-10, the observed EWs are considerably smaller than 
the maximum value predicted by the models, but
the relative line intensities are close to the predicted maximum.
This difference could be attributed to the shift between gaseous and 
stellar emission (see sect.~\ref{SPATA}). 

\subsection{Low-metallicity galaxies}

For the low-metallicity objects, the comparison between data and model 
predictions is much more complex. Only some of the observed WR line 
intensities and EWs lie well on the predicted model curves 
plotted in Fig.~\ref{MODELS}.
For the considered objects we dispose in total of 15 measurements 
of broad emission lines (all lines and all regions).
The relative line intensity of 7 WR regions  is above the maximum
value predicted by the standard model. On the other hand the majority 
(13 of 15) of the EWs are well in the range of the model 
predictions. The two remaining cases are $W($\heiia$)$ of
NGC~3125-A and Tol~89, which could well be overestimated due to nebular
contamination and contributions from other lines in the blended region,
and an exceptionally large \civ\ in Tol 89.
The uncertainties affecting the observed quantities are fully
discussed in SCK99.
Only for Tol~89 we have direct evidence that the displacement between gas and 
stars is likely of importance (see sect.~\ref{SPATA}).
For this galaxy we measure an extraordinarily strong \civ\ emission 
(see Table~\ref{DATA}) which exceeds the values shown in Fig.~\ref{MODELS}. 
We have no simple explanation for this strong emission. 
Other observations are required to confirm the measurements for 
Tol 89.

One possibility to explain the observations of some regions with 
large WR/\hbeta\ intensity ratios and large $W({\rm WR})$ may e.g.\ be 
a flatter IMF. Models with different burst durations and IMF have 
been tested in SCK99 (their fig. 12). 
These show that, within reasonable assumptions for the IMF slope,
the duration of star formation episode is limited to $\Delta t \la 2 - 4$ Myr.
What can be said about the IMF slope?
As mentioned earlier the observations of the WR features in NGC 3049, He 2-10,
and NGC 5253 are quite compatible with a Salpeter IMF. 
However, the large observed values of $W(4686)$ in NGC 3125-A and Tol 89, 
and $W(5808)$ in Tol 89 may require more WR stars. This can be obtained
by invoking a flatter IMF, which increases the WR equivalent widths.
However, we note that intrinsicly the model predictions for \heiia\ are not 
very sensitive to changes of the IMF slope.
More IMF-sensitive are the features of WC stars, which descend on the average 
from more massive predecessors than WN stars and from a narrower range of 
initial masses.
Although the standard models show some deficiencies, it is difficult to 
claim significant differences with respect to a Salpeter IMF (see SCK99 for 
details).
We therefore conclude that within the uncertainties our observations are 
compatible with a Salpeter IMF. 
Although we cannot exclude this possibility, no clear case requiring a 
significantly flatter IMF is found.
Much steeper IMF slopes may, however, not be compatible with our data.


\section{Conclusions}

The main result of this work is the unambiguous detection of
WC stars in five previously known WR galaxies, which brings to 
$\sim$ 54 the total number of extragalactic objects known 
to harbour both WN and WC stars (cf.\ SCP99).
The relative weakness of \civ\ compared to \heiia, and its larger width 
requires sufficiently high S/N ($\ga$ 40) to be detected. This explains most 
likely the non-detection in previous observations.

From the observed WR emission lines we concluded that all WR regions 
(except NGC~3049) contain a mixed population of WNL, and early WC and/or 
WO3-4 stars. Weak \ciii\ indicative of late WC stars is found in the 
high-metallicity WR region of NGC~3049. 
The estimated WC/WN number ratios (typically between 0.2 -- 0.4) show no 
clear trend with metallicity. 
For low-metallicity objects ($Z \sim 1/5$ \zsun), these values are larger than 
the observed WC/WN ratios in Local Group objects with similar $Z$ 
(MJ98). 
We argue that our WC/WN values are compatible with expectations for regions of 
short star formation. 
For He 2-10 and NGC 3049 the derived WC/WN ratio is below the trend given by 
MJ98. This can be due to the fact that we observe a short 
burst at a particular time (quite low probability), or that the number of WC 
stars is systematically underestimated. 
The solution awaits new observations and quantitative analysis of ``WC+WN 
galaxies'' of different metallicities.

Detecting both WN and WC features provides a considerable 
improvement to derive burst parameters. 
Given the WN $\rightarrow$ WC sequence, which is expected to be followed
only by the most massive WN stars, it is clear that the predictions of WC/WN 
populations are particularly sensitive to the evolutionary scenario and burst 
parameters (e.g.\ IMF, burst duration).

The comparisons of the observed EWs and line intensities 
relative to \hbeta\ with models do not all show a simple picture. 
The most important effects and uncertainties which may affect such a 
comparison have been amply discussed in SCK99.
In order to reproduce the WR lines, young populations with short durations 
of star formation are required. From our quantitative modeling, we find a
conservative limit for burst durations of typically $\Delta t \la 2 - 4$ Myr. 
The majority of the observed quantities can be reproduced reasonably 
well by the SV98 models with a Salpeter IMF. 
Although some WR lines may indicate a flatter IMF in some regions,
no clear case requiring a significantly flatter IMF is found.
Much steeper IMF slopes may, however, not be compatible with our data.
These results are in agreement with other studies of similar objects 
(see SCK99 and references therein).

\end{document}